\documentstyle[aps,prb,preprint,amsmath,tighten,graphicx]{revtex}
\begin{document}
\draft \title{Multi-terminal Molecular Wire Systems: A
Self-consistent Theory and Computer Simulations of Charging and
Transport}
\author{Eldon G. Emberly and George Kirczenow}
\address{Department of Physics, Simon Fraser University,
Burnaby, B.C., Canada V5A 1S6}
\date{\today}
\maketitle
\begin{abstract}
We present a self-consistent method for the evaluation of the
electronic current flowing through a multi-terminal molecular
wire. The method is based on B\"{u}ttiker-Landauer theory which
relates the current to one-electron scattering
probabilities. The scattering problem is solved using a
tight-binding form for Schr\"{o}dinger's equation that
incorporates a self-consistent evaluation of the electro-static
potential in the region of the molecular wire. We apply the
method to a three-terminal molecular wire connected to metallic
leads. The molecular wire is a $\pi$-conjugated carbon chain
with thiol end groups, self-assembled on the cleaved edge of a
multilayer of alternating thin metal and insulating films. The
ends of the chain bond to two outer metal layers that act as
source and drain, and the chain bridges a third (inner) metal
layer that acts as a gate. We show that transistor action
should occur in this device if on the surface of the metal gate
there are adsorbed atoms that acquire charge as the gate
voltage is increased, thereby enhancing the interaction between
the gate and molecule and creating a strong potential barrier
that hinders electron flow along the molecular wire. We find
that electronic solitons play an important role in the response
of this system to applied voltages.
\end{abstract}



\section{Introduction}
A molecular wire in its simplest definition is a single
molecule which is used as a bridge to carry electronic current
between metallic contacts. Recently developed techniques that
allow manipulation of matter at the atomic scale have made
remarkable experiments on molecular wires possible.  Some of
the first experiments measuring the current through single
molecules employed the scanning tunnelling microscope
(STM)\cite{Andres96,Bumm96} or used self-assembly to bridge the
gap between two nanoscopic metal leads with a single molecule
(1,4 benzene-dithiolate) whose current-voltage characteristic
was then measured.\cite{Reed97} These and other two terminal
measurements\cite{Reed99,Gim99,Stipe98,Joach97} have shown that
a molecular wire does not behave like a simple ohmic resistor
but displays features arising from the quantum states of the
molecule.  Theoretical work on molecular wires has proposed
various models for the explanation of the observed phenomena
and has until now focused on two-terminal
devices.\cite{Datta97,Yaliraki99,Lang00,DiVentra00,Emberly98}

For two terminal mesoscopic systems, Landauer
theory\cite{Lan57} relates the electronic current to the
transmission probability for a single electron to scatter
through the system.  Coherent transport in multi-terminal
mesoscopic systems has been treated by B\"{u}ttiker who has
given an extension of Landauer theory\cite{Buettiker}, and
Christen and B\"{u}ttiker have discussed charging effects in
such systems.\cite{Christen96} However, to date there has been
very little theoretical work done addressing these issues in
the context of multi-terminal molecular wire
systems.\cite{Emberly00}

In this paper, we present a self-consistent method based on
B\"{u}ttiker-Landauer theory for the evaluation of electronic
current in a multi-terminal molecular wire system. This method
considers explicitly the effects of charging in molecular wires
due to the application of voltages to the terminals.  The
theory is then applied to a molecular wire transistor (a
three-terminal molecular wire device). The system we consider
is a $\pi$-conjugated chain molecule with thiol end groups
self-assembled onto the cleaved edge of a multilayer
substrate. The substrate consists of alternating metal and
insulating thin films. The chain bridges two of the metal films
and passes over a third metal film which acts as a gate. For
the system considered we find that transistor-like behavior is
possible if the electric field of the gate is enhanced by
chargeable adsorbates on its surface.  We also find that charge
density waves and charged solitons form in this molecular wire
system and play an important role in its response to applied
bias voltages.

Sec.~I summarizes briefly the necessary results from
B\"{u}ttiker-Landauer theory for the evaluation of the
electronic current in a multi-terminal molecular wire. A method
for solving the single-electron scattering problem is presented
in Sec.~II. Our self-consistent method for determining the
electro-static potential is presented in Sec.~III. These
methods are then applied to the calculation of the charging
behavior and current vs.~gate bias voltage characteristics of a
three terminal molecular wire transistor in Sec.~IV. Our
conclusions are given in Sec.~V. An appendix describing the
electrostatics of the molecular wire configuration that was
studied is also included.

\section{Multi-terminal B\"{u}ttiker-Landauer Theory}
For two-terminal mesoscopic systems (such as a molecule
bridging two metallic electrodes), the electronic current can
be calculated using Landauer theory.\cite{Lan57} The systems
that will be of interest in this article have more than two
metallic contacts and thus it is necessary to have a method for
calculating the current in these multi-terminal
structures. B\"{u}ttiker and Christen have generalized
Landauer theory to provide a theoretical framework
for treating transport at finite voltages for systems connected
to more than two leads.\cite{Buettiker,Christen96} We give a
brief summary of their relevant results.

Consider a mesoscopic conductor (such as a molecular wire)
attached to $N$ different terminals or reservoirs $\alpha =
1\ldots N$.  Each of these reservoirs may be at a different
electro-chemical potential $\mu_\alpha$. The important quantity
is the difference between each of these electro-chemical
potentials and a common reference electro-chemical potential
$\mu_0$, $\delta \mu_\alpha = \mu_\alpha - \mu_0$. This can be
related to the bias voltage applied to the reservoir by
$\delta\mu_\alpha = - e V_\alpha$. The current flowing into one
of these reservoirs will be related to the probability for an
electron to scatter into that reservoir, however now there
are more scattering processes that can occur. An electron
scattering from the $\alpha'$ reservoir to the $\alpha$
reservoir is assigned a transmission probability
$T_{\alpha,\alpha'}$. Reflection within the $\alpha$ reservoir
is described by the reflection probability $R_\alpha$. The net
current flowing into the $\alpha^{th}$ reservoir has several
contributions. Electrons scatter into it from every other
reservoir, and the current contributed by the $\alpha'$
reservoir is $\delta i_{\alpha,\alpha'} = -(2 e/h)
T_{\alpha,\alpha'}\delta\mu_{\alpha'}$. There is also a
contribution due to reflected electrons and this is given by
$\delta i_{\alpha,\alpha} = -(2 e/h) R_\alpha
\delta\mu_\alpha$. Adding these two contributions gives the
total current flowing into reservoir $\alpha$. However there is
also current flowing out of reservoir $\alpha$ and this is
given by $-(2 e/h) \delta\mu_\alpha$. The net current flowing
into reservoir $\alpha$ is given by the difference between
these contributions. Combining these terms the current
contribution in the lead connected to terminal $\alpha$ due to
small differences in the electro-chemical potentials is given
by
\begin{equation}
\delta i_\alpha = -\frac{2 e}{h} \{(1 - R_\alpha)\delta\mu_\alpha -
\sum_{\alpha'} T_{\alpha,\alpha'} \delta\mu_{\alpha'} \}.
\end{equation}
This can be generalized to finite temperature $T$ and finite
voltages by integrating over the Fermi distributions,
$F(E,V_\alpha)=1/(\exp[-(E-\mu_\alpha)/kT]+1)$, of each of the
reservoirs. This gives,
\begin{equation}
i_\alpha = -\frac{2 e}{h} \int dE \{(1-R_\alpha(E,\{V_\gamma\})
F(E,V_\alpha) - \sum_{\alpha'} T_{\alpha,\alpha'}(E,\{V_\gamma\})
F(E,V_{\alpha'}) \}. \label{eq:multi-landauer}
\end{equation}
This equation properly reduces to the two terminal result if
$T_{1,2} = T_{2,1} = T$ and $1 - R_{\alpha} = T$.

The transmission and reflection probabilities that enter
Eq.~(\ref{eq:multi-landauer}) depend on the electric fields
throughout the system that arise from the application of the
bias voltages and from the charges on the molecule.  A later
section provides a self-consistent method for handling this
problem.

\section{Multi-terminal Scattering Theory}
A variety of techniques exist for the evaluation of the
relevant scattering probabilities that go into the calculation
of the current. The method that we use determines these
quantities directly from the single electron wave function.
The wave function is calculated by solving Schroedinger's
equation for the scattered electron. Thus the problem is to
find the wave function $|\Psi^\alpha\rangle$ for an electron
with energy $E$ incident from reservoir $\alpha$. This wave
function satisfies
\begin{equation}
[H - e\Phi(\{q\},\{V\})]|\Psi^\alpha\rangle
= E|\Psi^\alpha\rangle
\end{equation}
where $H$ is the single electron Hamiltonian of the coupled
system in the absence of applied biases and $\Phi$ is the
electro-static potential that depends upon the voltages $\{V\}$
applied to the terminals and the charges $\{q\}$ in the
system. For simplicity, we assume that each of the reservoirs
has only one electronic mode (the generalization to
multi-channel reservoirs is straightforward but the notation
becomes complicated). We represent the system of leads and
molecule using the tight-binding approximation with an
orthogonal set of atomic orbitals. The molecule has a set of
atomic orbitals $\{|\phi_j\rangle\}$. Each of the $N$
reservoirs also has a set of orbitals and these are denoted
$\{|n_\alpha\rangle\}$ where $n_\alpha = 1\ldots\infty$ on the
$\alpha^{th}$ reservoir. The complete wave function can be
written as a sum of the wave functions in each of the
reservoirs $\alpha$ and the wave function on the molecule (M)
$|\Psi^\alpha\rangle = \sum_{\alpha'} |\psi_{\alpha'}\rangle +
|\psi_M\rangle$. The boundary conditions on these wave
functions are the following for an electron incident from the
$\alpha^{th}$ reservoir:
\begin{subequations}
\begin{align}
|\psi_{\alpha}\rangle &= \sum_{n=1}^{\infty} \exp(-i n
  y^{\alpha})|n_\alpha\rangle + r_\alpha \sum_{n=1}^{\infty} \exp(i n
  y^{\alpha}) |n_\alpha\rangle \\ |\psi_{\alpha'}\rangle &=
  t_{\alpha',\alpha} \sum_{n=1}^{\infty} \exp(i n y^{\alpha'})
  |n_{\alpha'}\rangle \\ |\psi_M\rangle &= \sum_j c_j^\alpha
  |\phi_j\rangle
\end{align}
\end{subequations}
Inserting the above wave function into Schroedinger's equation
yields an infinite set of linear equations in the unknowns
$r_\alpha$, $t_{\alpha',\alpha}$ and $c_j^\alpha$. Applying the
bras $\langle 1_\alpha|$ where $\alpha=1\ldots N$ and
$\{\langle \phi_j|\}$ yields a solvable set of linearly
independent equations.

With the scattering coefficients determined the transmission
and reflection probabilities for multi-terminal scattering are
found from
\begin{subequations}
\begin{align}
T_{\alpha',\alpha} &= \frac{v_\alpha'}{v_\alpha} |t_{\alpha',\alpha}|^2
\\
R_{\alpha} &= |r_\alpha|^2
\end{align}
\end{subequations}
where $v_\alpha$ is the velocity of an electron with energy $E$
in the $\alpha^{th}$ reservoir.

\section[Non-equilibrium Charge Distribution]{Non-equilibrium Charge 
Distribution and the Self-consistent Potential} 
In this section the question as to how to deal with the
electric field due to the applied voltages and the charge in
the system will be addressed.

The electro-static potential energy $-e \Phi$ has several
contributions. First, there is the contribution due to the
electro-static potential due to the presence of the reservoirs,
each at a different voltage bias. This is found from the
solution to Laplace's equation $\nabla^2
\phi(\vec{r},\{V_\alpha\}) = 0$ subject to the boundary
condition that the surface of each reservoir is an
equipotential . Secondly, the electro-static field will induce
a non-equilibrium charge distribution on the molecule. Because
of this and because of chemical charge transfer effects between
different atoms that occur even in the absence of bias
voltages, the atoms along the wire will no longer be
electrically neutral, and this can be quantified as fractional
ionization charge $q_j$ for atom $j$ on the molecule ($q_j < 0$
corresponds to a negatively charged ion). At the mean-field
level of approximation a propagating electron at position
$\vec{r}$ will now interact with these charges via an
unscreened Coulomb potential. And lastly, these induced charges
will in turn generate image charges in all of the metallic
contacts. Thus the electrostatic energy terms that must be
included in a mean field treatment of the molecular wire are
\begin{equation}
-e \Phi = -e \phi(\vec{r},\{V_\alpha\}) - e \sum_j \frac{q_j}{4 \pi
\epsilon_o}\frac{1}{|\vec{r}-\vec{r}_j|} - e \phi_{image}(\vec{r},\{q_j\}).
\end{equation}
The potentials $\phi(\vec{r},\{V_\alpha\})$ and
$\phi_{image}(\vec{r},\{q_j\})$ both depend on the system's
geometry. A solution for the specific three terminal geometry
considered below is presented in the Appendix.

Within TBA, the presence of the electrostatic terms modifies
the site
energy of each orbital on the molecule. For instance, if
each atom of the molecule only has a single orbital, then the
diagonal element $H_{i,i} = \langle i|H-e\Phi| i\rangle$ for
the $i^{th}$ atom of the molecule becomes
\[
H_{i,i} = \epsilon_i(q_i) - e \phi(\vec{r}_i) - e \sum_{j\ne
i}\frac{q_j}{4 \pi \epsilon_o}\frac{1}{|\vec{r}_i-\vec{r}_j|} - e
\phi_{image}(\vec{r}_i,\{q_j\})
\]
where $\epsilon_i(q_i)$ is a charge dependent site energy for
atom $i$. In the chemistry literature $\epsilon_i(q_i)$
is usually taken
to be a linear or quadratic fit to ionization potential energy
data. Thus
\[
\epsilon_i(q_i) = \epsilon_i^0 + q_i U + (\mathrm{term\: in\: } q^2)
\]
where $\epsilon_i^0$ is roughly equal to the average of the
ionization energies for the +ve ion and -ve ion (although the
best fit may depart from this), and $U$ is an energy associated
with fractional charging.\cite{Mcglynn72} This treatment of the
on-site energies can also be regarded as a tight-binding
version of the local density approximation since the first
order variation of the charge density contribution to the site
energy generates a linear term associated with the derivative
of the site energy with respect to the fractional charge.

Thus the problem that now needs to be addressed is the
determination of the charge on the wire. This is a
self-consistent problem. The scattering problem must be solved
to determine the charge. Then this charge must be used to
adjust the effective Hamiltonian, and then the scattering
problem must be solved again. This must be repeated until the
charge on the wire converges.

In each iteration the charge is calculated as follows: Since
the wave function is being calculated explicitly from
Schroedinger's equation, the probability of an electron being
on a certain site $j$ on the molecule is given by $|c_j|^2$ for
a given scattering state. The spectral charge on that site
contributed by an electron incident with energy $E$
(propagating in a Bloch state with wave vector $ k^\alpha$) from
reservoir $\alpha$ is simply $-e |c_j^\alpha|^2$. Hence the
charge on site $j$ due to electrons incident from lead $\alpha$
is found by integrating over all the occupied Bloch states in
that reservoir. The total charge is then found by summing over
all the reservoirs. [Normalization has not been considered
yet. The proper normalization for a scattering state
$|\Psi^\alpha\rangle$ is $1/\sqrt{N}$ (where $N$ is the number
of unit cells) since this is the normalization for a
propagating Bloch wave that will exist deep within the
reservoirs. Thus in the above, the charge needs to be divided
by $N$]. Thus the charge can be written (including a factor
of 2 for spin) as
\[
q_j = - 2 e \sum_{\alpha} \sum_{k^{\alpha}}
|\Psi_j^{\alpha}(k^\alpha)|^2 = - \frac{2 e}{N} \sum_{\alpha}
\sum_{k^{\alpha}} |c_j^{\alpha}(k^\alpha)|^2
\]
Converting the wave vector sum into an integral (and making a
change of variables to the reduced wave vector $y^\alpha =
k^\alpha a$) results in
\[
q_j = -\frac{2 e}{N} \frac{L}{2\pi a}\sum_{\alpha} \int_0^{y_F} dy^\alpha\;
|c_j^{\alpha}(y^\alpha)|^2
\]
Finally, converting the above wave vector integral into an
energy integral over the energy band of each reservoir
$\alpha$, using $L = N a$, and generalizing to finite
temperatures and voltages by using the Fermi function for
reservoir $\alpha$, $F(E,V_\alpha)$, the following expression
is arrived at for the total charge on atom $j$ of the molecule,
\begin{equation}
q_j = -\frac{2 e}{2 \pi} \sum_{\alpha} \int dE\;
\frac{dy^\alpha}{dE}|c_j^{\alpha}(E)|^2 F(E,V_\alpha)
\end{equation}
We now consider an application of the above method to a
multi-terminal molecular wire system.

\section{Molecular Transistor}
In this section we present a feasible suggestion for a
molecular wire transistor (MWT) and study it theoretically by
the methods described above. A simple transistor functions by
transmitting a current between an electron source and drain and
tuning this current by the application of a voltage to a
gate. At low gate voltages, current flows freely through the
transistor under an applied bias between the source and
drain. As the gate voltage increases an electro-static barrier
forms and the current decreases significantly. Whether behavior
of this type can be exhibited by a single-molecule device (a
MWT) has been a long-standing open question both experimentally
and theoretically: The solid state transistor operates
essentially on classical electro-magnetic principles -- an
electric field suppresses the flow of electrons which is
governed by Boltzmann diffusion. When electrons flow through a
molecule, their dynamics is governed not by classical physics
but by quantum physics and the length and energy scales
involved are also very different from those in solid state
transistors. We show that these physical differences are very
important, but that they can be addressed and that a working
MWT should be feasible.

\begin{figure}[!t]
\begin{center}
\includegraphics[bb = 0 0  640 800,clip,width =
0.75\textwidth]{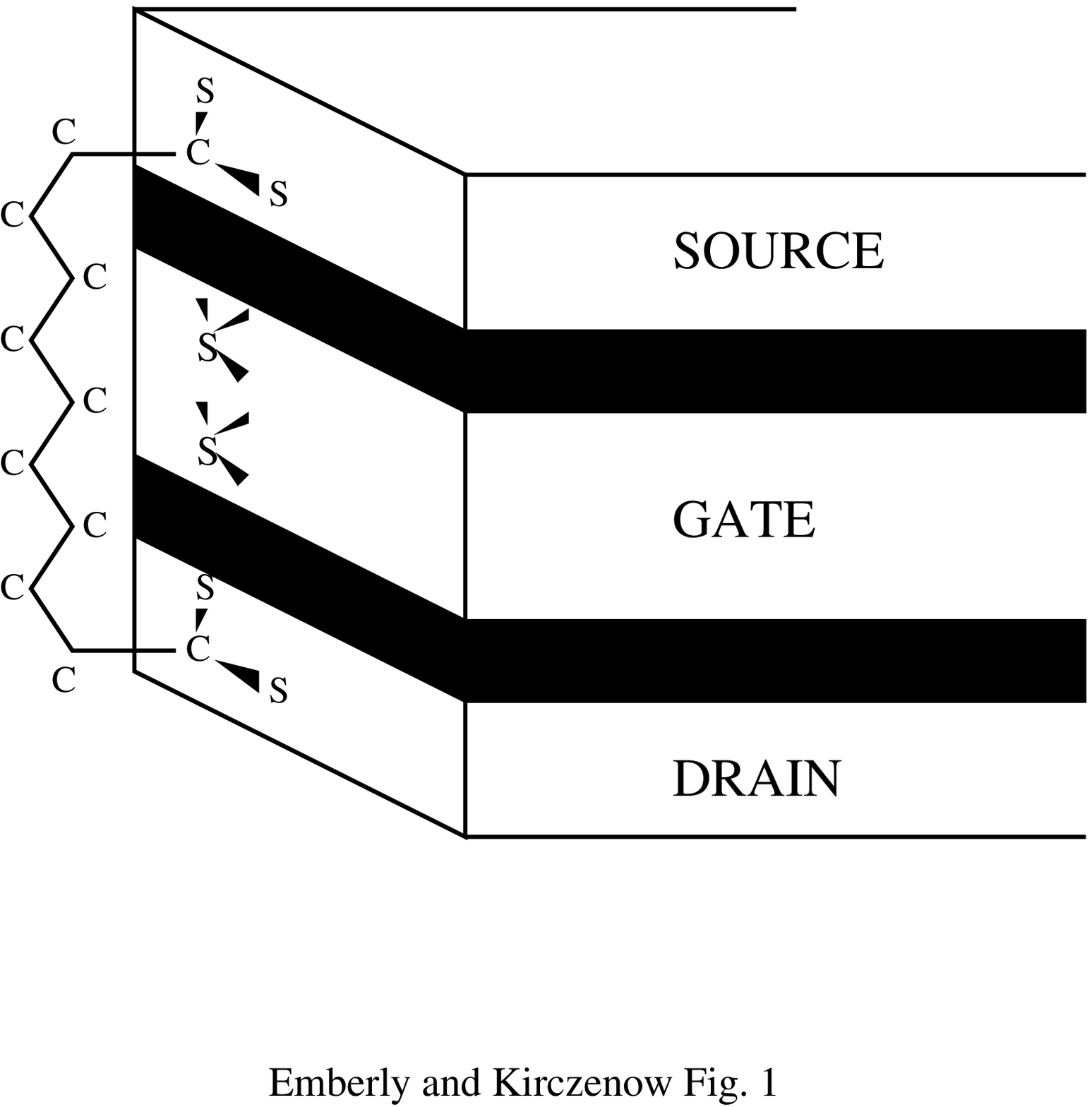}
\caption[Schematic of a molecular wire transistor]{Schematic of a
molecular wire transistor. The substrate contains three gold layers
that act as source, drain and gate. The source is bridged to the drain
by a $\pi$-conjugated carbon chain. Sulfur ions on the gate enhance
the electro-static coupling between the gate and carbon chain.}
\label{fig1}
\end{center}
\end{figure}

The greatest experimental challenge in developing a MWT is the
introduction of the third terminal which should act as the
gate.  To date this has been achieved only in experiments on
carbon nanotubes that were contacted lithographically to metal
leads at the ends and then a gate potential was applied via an
STM tip that was brought close to the middle of the
tube.\cite{Tans97,Tans98} In the present work
a three terminal structure
consisting of a molecule {\em self-assembled} on the cleaved
edge of a multilayer consisting of three metallic films each a
few monolayers thick and separated from the other metal layers
by thin insulating layers (as is shown schematically in
Fig.~\ref{fig1}) is considered. In the figure the outermost
metal layers act as source and drain, and the inner layer is
the gate. Although such structures have not yet been realized
experimentally, it is reasonable to expect that they will be
fabricated in the future. Our purpose here is to explore
theoretically the fundamental principles that will help to
guide experimental work on such systems.

The molecule that is considered is a $\pi$-conjugated carbon
chain molecule terminated with thiol (sulfur) groups. The
$\pi$-conjugation provides for relatively easy electron
transport along the chain between source and drain; such chains
when strongly coupled to metallic leads are very conductive
(relative to other molecular conductors). The thiols make for
strong bonds between the ends of the chain and the source and
drain respectively and are responsible for self-assembly of the
MWT on the substrate. In the absence of the gate, this is just
a two terminal molecular device, broadly similar to the
molecular wires that have already been discussed and studied
experimentally.\cite{Andres96,Bumm96,Reed97}

The conducting layers of the substrate are assumed to be
gold. At the Fermi energy of gold, electrons on the molecular
chain reside in orbitals that are $\pi$ like. Thus when the
chain is bonded between the source and drain, transport of
electrons is along the conjugated $\pi$ backbone of the
chain. In the self-assembled structure shown in
Fig.~\ref{fig1} the carbon chain is oriented such that these
$\pi$ orbitals are parallel to the substrate's surface; this
implies a low probability of electron hopping between the
surface and molecule.

In the tight-binding scheme, only the $\pi$ orbital of each
carbon atom is taken into account and it is assigned a site
energy of $-11.4$~eV. The structure of the chain is taken to be
undimerized trans-polyacetylene. Only nearest neighbour hopping
is considered on the chain and a value of $-2.5$~eV is assigned
to the hopping matrix element. Transport in the gold terminals
in the substrate of Fig.~\ref{fig1} is modeled using single
channel ideal leads. Since gold is an s-orbital metal, only its
6s orbital is considered and this is assigned a site energy of
$-10$~eV (the equilibrium Fermi energy for this model). The
hopping energy of the gold leads was taken to be $-3.5$~eV. The
sulfurs are modelled using their 3p orbitals which have a site
energy of $-11$~eV.  To simulate charge transfer, a single gold
atom is bonded beneath each sulfur atom. (If the sulfur atoms
were coupled directly to the ideal leads, the surface charging
that occurs at the interface between the sulfur atoms and the
gold surface would be neglected. These single gold atoms are
used to simulate the physics of the surface layer of gold and
the chemistry of the sulfurs binding over hollow sites on the
gold surface). Each of these gold atoms is coupled to a single
channel lead. They are positioned 2~\AA~ above the
equipotential surface (depicted in Fig.~\ref{fig8} of the
Appendix). Thus the equipotential surface is assumed to begin
below the surface layer of the gold leads. To simulate the
strong binding that occurs when sulfur binds over a hollow site
on the gold surface, each of these gold atoms is strongly
bonded to its sulfur atom and this is quantified by a hopping
energy of $-2$~eV. Each sulfur is assumed to be positioned
2~\AA~ above its gold atom. Similarly the binding between the
sulfurs and the carbon is assumed to be strong, and this
hopping energy is assigned a value of $-2$~eV. The
sulfur-carbon bond is taken to be 1.8~\AA. The fractional
charging energy $U$ that multiplies the linear term in $q$ in
the site energy for each of these orbitals was taken to be
11.5~eV for the C 2p, 9.7~eV for the S 3p, and 8.6~eV for the
Au 6s.\cite{alvarez85}

\begin{figure}[!t]
\begin{center}
\includegraphics[bb = 0 0  640 800,clip,width =
0.75\textwidth]{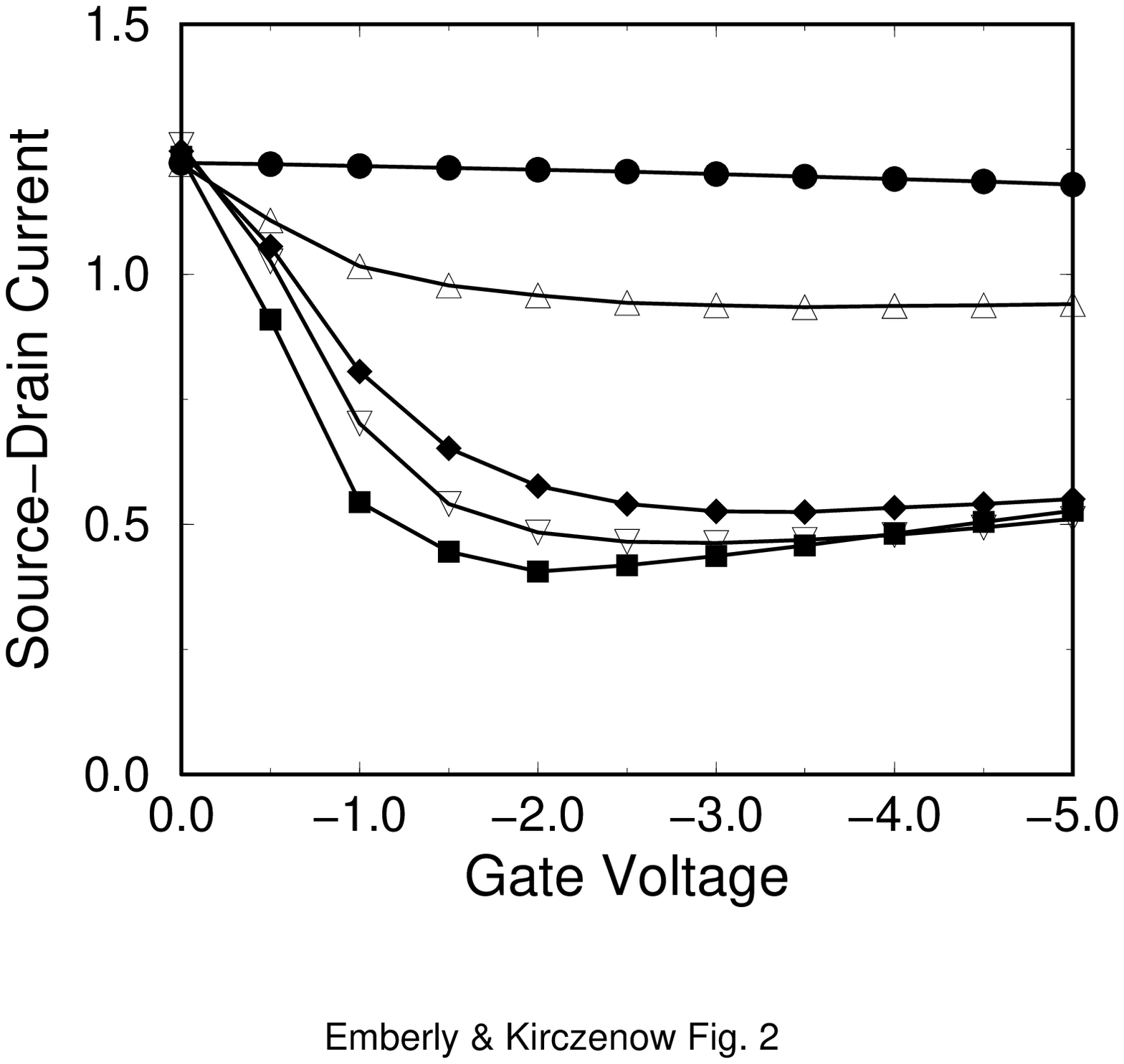}
\caption[The current as a function of gate bias voltage]{The current
as a function of gate bias voltage. The source-drain bias is
1~V. Circles: $C_{13}S_4H_{11}$ with no S atoms on the gate. Up triangles:
$C_{13}S_4H_{11}$ with one S on the gate. Diamonds: $C_{15}S_4H_{13}$
with two S on the gate. Down triangles: $C_{17}S_4H_{15}$ with three S
on the gate. Squares: $C_{19}S_4H_{17}$ with four S on the
gate. The units of current are $(2e/h)~eV$.}
\label{fig2}
\end{center}
\end{figure}

The above methods together with the results of the Appendix for
the electrostatic potential due to the metal terminals were
used to calculate the current as a function of gate voltage
through a MWT based on $C_{13}S_{4}H_{11}$ molecule. (This is
similar to the molecule depicted in Fig.~\ref{fig1} but has two
more $CH$ groups). A symmetric source-drain bias voltage of 1~V
was assumed and gate voltages up to a bias of -5~V (i.e., +5~eV
for an electron) were studied. We found that charge is
transferred to the chain. This is consistent with the polymer
literature which reports charge transfer from the metal to
carbon chains. However it was also found that the electric
current along the carbon chain was resistant to the application
of the gate voltage. Electron transport along the backbone remained
unhindered, and the current was almost independent of the gate
bias voltage (the circles in Fig.~\ref{fig2}). This indicates
that in a functioning MWT the gate needs to interact with the
chain molecule much more strongly than it does in the system
considered above.

Our proposed solution to this problem is to adsorb onto the
surface of the gate an atomic species that charges up as the
gate voltage is applied and thus enhances the effective
electrostatic interaction between the gate and chain
molecule. We have chosen sulfur since it binds readily to gold
surfaces and gains charge from the gold surface of the gate as
a bias is applied to the gate in this model.  (For the narrow
nanoscale gates considered here, there is a substantial
electric field in the vicinity of the gate due to the applied
gate bias which helps to push charge onto the adsorbed sulfur
atom. For much wider gates this effect is lessened
significantly as the surface electric field is much
weaker). For the same chain molecule as above, a single sulfur
atom adsorbed on the gate was modeled by situating it
2.5~\AA~below the middle carbon atom of the chain. Again this
sulfur atom was bonded over a single gold atom so as to
simulate the charge transfer that would take place in the
physical system. The gold atom was then attached to
a single-channel ideal lead. Now as the gate voltage is
applied, the sulfur charges up, and this extra local charge
results in an electro-static potential barrier that can
significantly reduce the current flowing through the molecular
wire.  [Note that this sulfur atom should not bond to the $\pi$
carbon backbone since its p orbital is orthogonal to
it.\cite{sellers93} Although it may modify the $\sigma$ bonding
structure, it should also not form any $\sigma$ bonds since
these are saturated along the carbon chain.] The results are
shown in Fig.~\ref{fig2} (up triangles) for a 1 V source-drain
bias. There is now a clear reduction in the current as the gate
voltage is increased.

A MWT with a second sulfur atom also adsorbed on the gate is
depicted in Fig.~\ref{fig1}. (The sulfur atoms on the gate
are 2.8~\AA~apart -- this corresponds roughly to the distance
between hollow sites on an ideal (100) gold surface). The two
sulfurs together have a stronger effect on the current. This is
shown in Fig.~\ref{fig2} (diamonds); the current is now
reduced by a factor of 3 at a gate voltage of -2.5 V.  This
significant reduction in the current is due to the formation of
a large potential energy barrier along the wire.

\begin{figure}[!t]
\begin{center}
\includegraphics[bb = 0 0  640 800,clip,width =
0.75\textwidth]{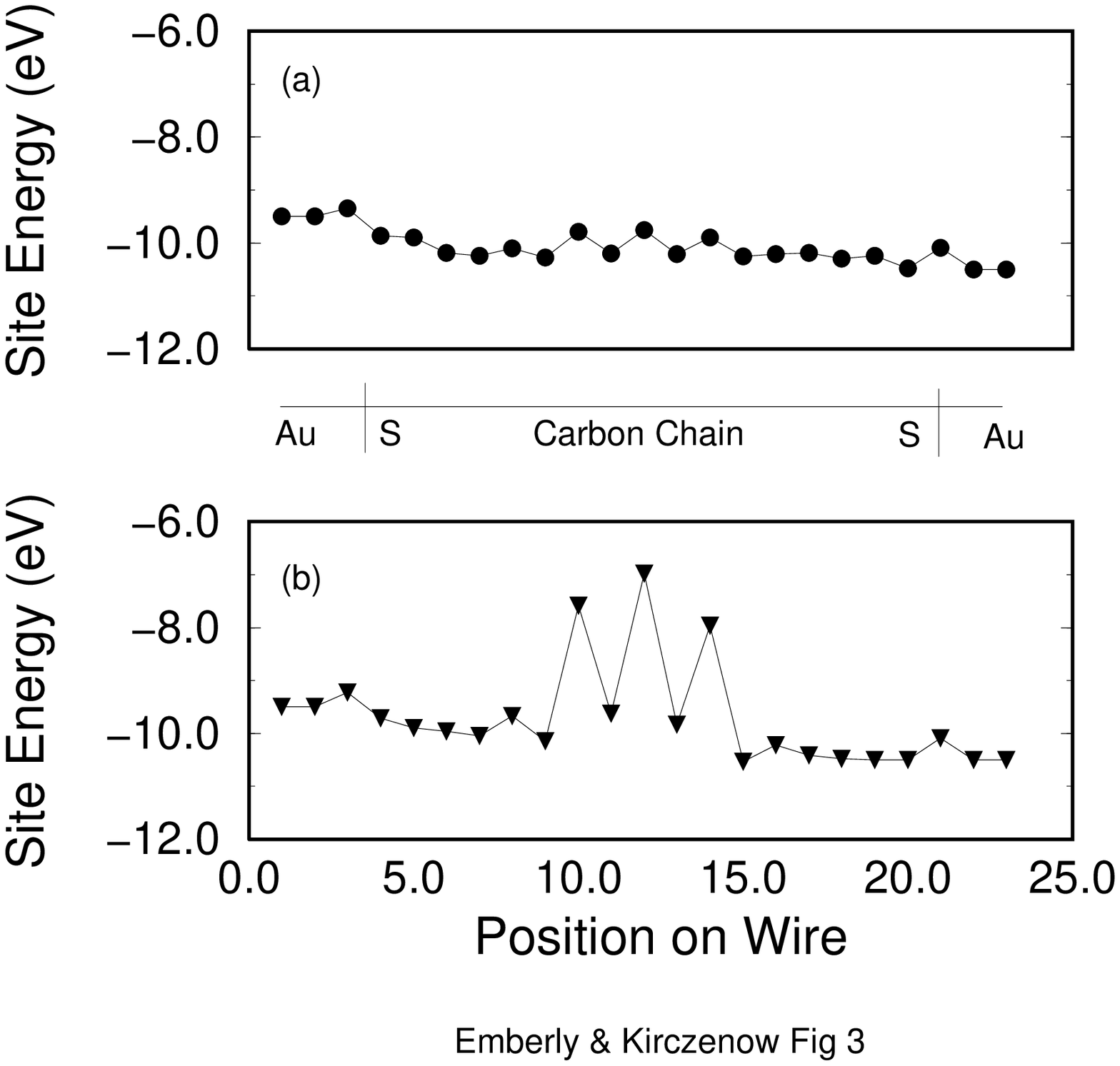}
\caption[Site energy of each atom along
the chain]{This figure depicts the site energy of each atom along the
chain. For each graph, the source-drain bias voltage was 1 V. (a)
  Graph for gate bias of 0 V and (b) graph for a gate bias of -3 V.}
\label{fig3}
\end{center}
\end{figure}

The self-consistently calculated site energy of each atom is
plotted in Fig.~\ref{fig3}.  First the 0 V gate voltage result
(Fig.~\ref{fig3}a) should be noted. The source is held at
$-0.5$~V (+0.5~eV electron energy) and the drain at
+0.5~V. Right at the interface between the gold and the thiol,
a potential barrier is formed due to charge
transfer. (Including the single gold atoms between the ideal
leads, which are assumed to form an equipotential, and the
sulfur allows this barrier to be modeled in the present
calculations).  The potential along the carbon chain is roughly
flat due to charge re-arrangement on the
chain. Fig.~\ref{fig3}b shows the site energy along the wire at
a gate voltage of $-3$~V. There is now a significant
electrostatic potential barrier in the middle of the carbon
chain. The potential barrier is not smooth, but consists of
peaks and valleys due to alternating distances between the
carbon atoms and the adsorbed sulfur atoms on the gate in
Fig.~\ref{fig1}.

\begin{figure}[!t]
\begin{center}
\includegraphics[bb = 0 0  640 800,clip,width =
0.75\textwidth]{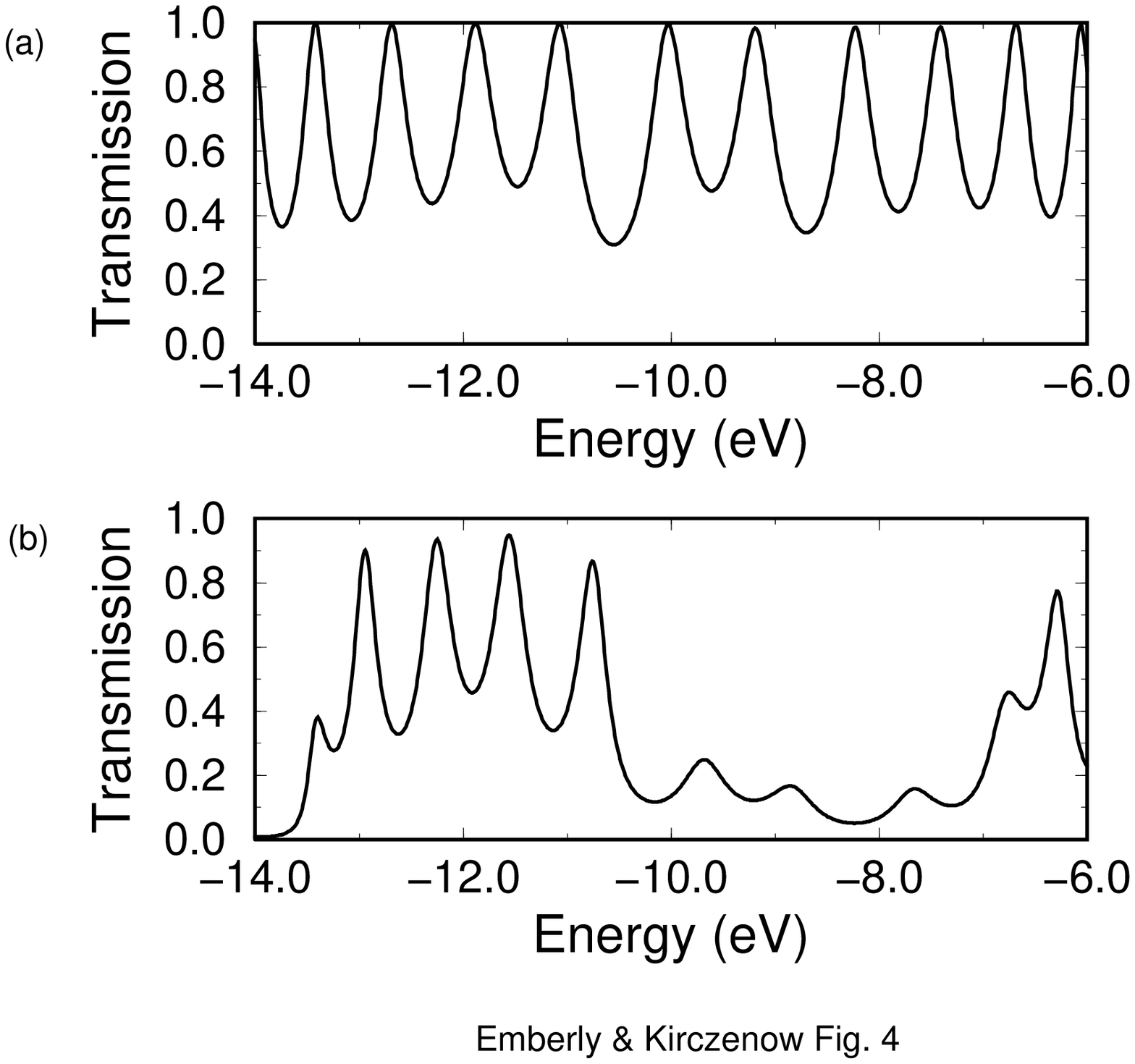}
\caption[Transmission probability for an electron to scatter
through the chain]{Transmission probability for an electron to scatter
through the chain. (a) Corresponds to a gate bias of 0 V
and (b) has a gate bias of $-5$~V. The source and
drain are biassed at 1 V.}
\label{fig4}
\end{center}
\end{figure}

This barrier hinders electron transport as is seen in the plot
(Fig.~\ref{fig4}) of the transmission probability of an
electron from the source to the drain.  The reduction in
transmission probability is associated with the formation of a
gap in the sequence of resonant states around the Fermi energy,
as can be clearly seen in Fig.~\ref{fig4}: At zero gate voltage
(Fig.~\ref{fig4}a) there is a strong resonance at the Fermi
energy of $-10$~eV. Fig.~\ref{fig4}b corresponds to a gate
voltage of $-3$ V and there is now a clear gap between the
resonant states around the Fermi energy. For longer carbon
chains with more sulfurs on the gate the nearly periodic
structure of the barrier creates the equivalent of a band gap
with the Fermi energy lying near the edge of the lower energy
band.  The molecule transmits electrons quite well at energies
outside of the band gap that forms around the Fermi energy.
Increasing the length of the barrier suppresses transmission
further in the band gap. However because the Fermi energy lies
near the edge of the bottom transmitting band, as the gate
voltage is increased further, eventually the resonant states in
the lower band begin to transmit a net current and the current
rises. Stronger transistor action could be achieved for a Fermi
energy more centered within the gap. However, in the
simulations, charge transfer has the effect of keeping the
lower lying resonances fairly close to the Fermi energy. Up to
4 sulfur atoms on the gate were simulated, thus increasing the
length of the barrier and further reducing the current as shown
in Fig.~\ref{fig2}. For these longer chains the current rises
at higher gate voltages as expected from the above arguments.

\begin{figure}[!t]
\begin{center}
\includegraphics[bb = 0 0  640 800,clip,width =
0.75\textwidth]{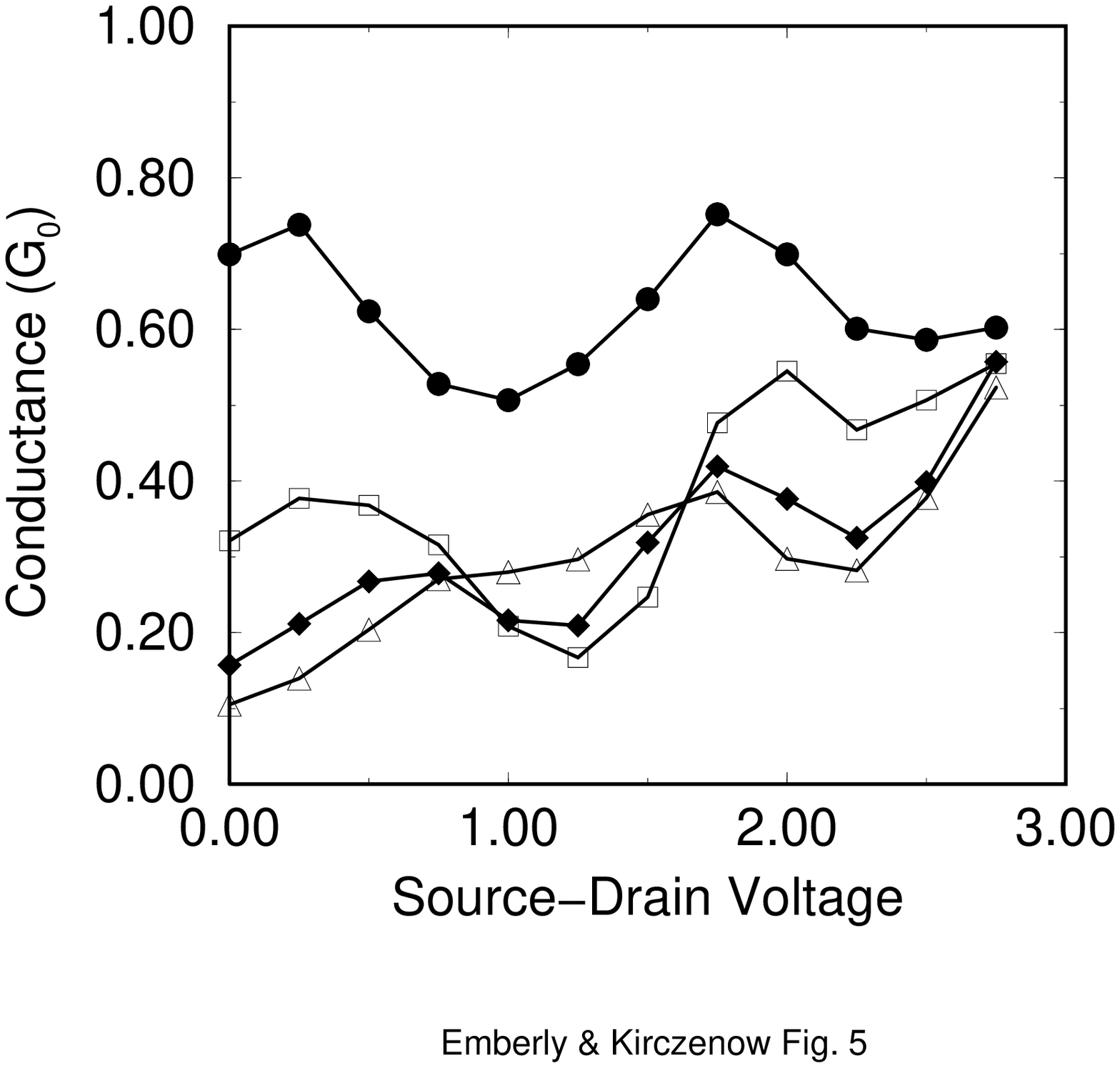}
\caption{Differential conductance of molecular wire as a function of
gate bias and source-drain bias: (filled circles) zero gate bias,
(open squares) 1 V gate bias, (filled diamonds) 2 V gate bias, (open
triangles) 3 V gate bias. The units are in $G_0 = 2 e^2/h$.}
\label{fig5}
\end{center}
\end{figure}

To gain further insight into the transport properties of these
chains, the differential conductance of the molecular wire with
two sulfurs adsorbed on the gate is plotted at various gate
voltages as a function of source-drain bias in Fig. \ref{fig5}.
The figure shows at which source-drain bias and at which gate
voltage optimal transistor action can be achieved. The top curve
with filled circles corresponds to zero gate bias. The peaks in
the conductance correspond to resonant transmission through the
resonances near the Fermi energy. The other graphs correspond
to different reverse biases applied to the gate. For low
source-drain voltages, the conductance of the molecule is
clearly lowered due to the barrier being formed. However
because the Fermi energy of the leads resides near the edge of
the lower band of resonant states as was discussed above, for
sufficiently high source-drain biases, the molecule can be made
to conduct appreciably even when a large reverse bias is
applied to the gate. Optimal transistor action can be achieved
by operating in a regime where there are clearly high and low
conducting states. This can be achieved at source-drain biases
near 0 V by varying the gate bias from $0$ V to $-3$ V. Here
the conductance varies from approximately 0.7 $G_0$ to 0.1
$G_0$. Other favourable operating voltages correspond to a
source-drain bias around 1 V and a gate bias between $-1$ and
$-2$ V. Here the conductance varies between 0.5 $G_0$ and 0.2
$G_0$.

\begin{figure}[!t]
\begin{center}
\includegraphics[bb = 0 0  640 800,clip,width =
0.75\textwidth]{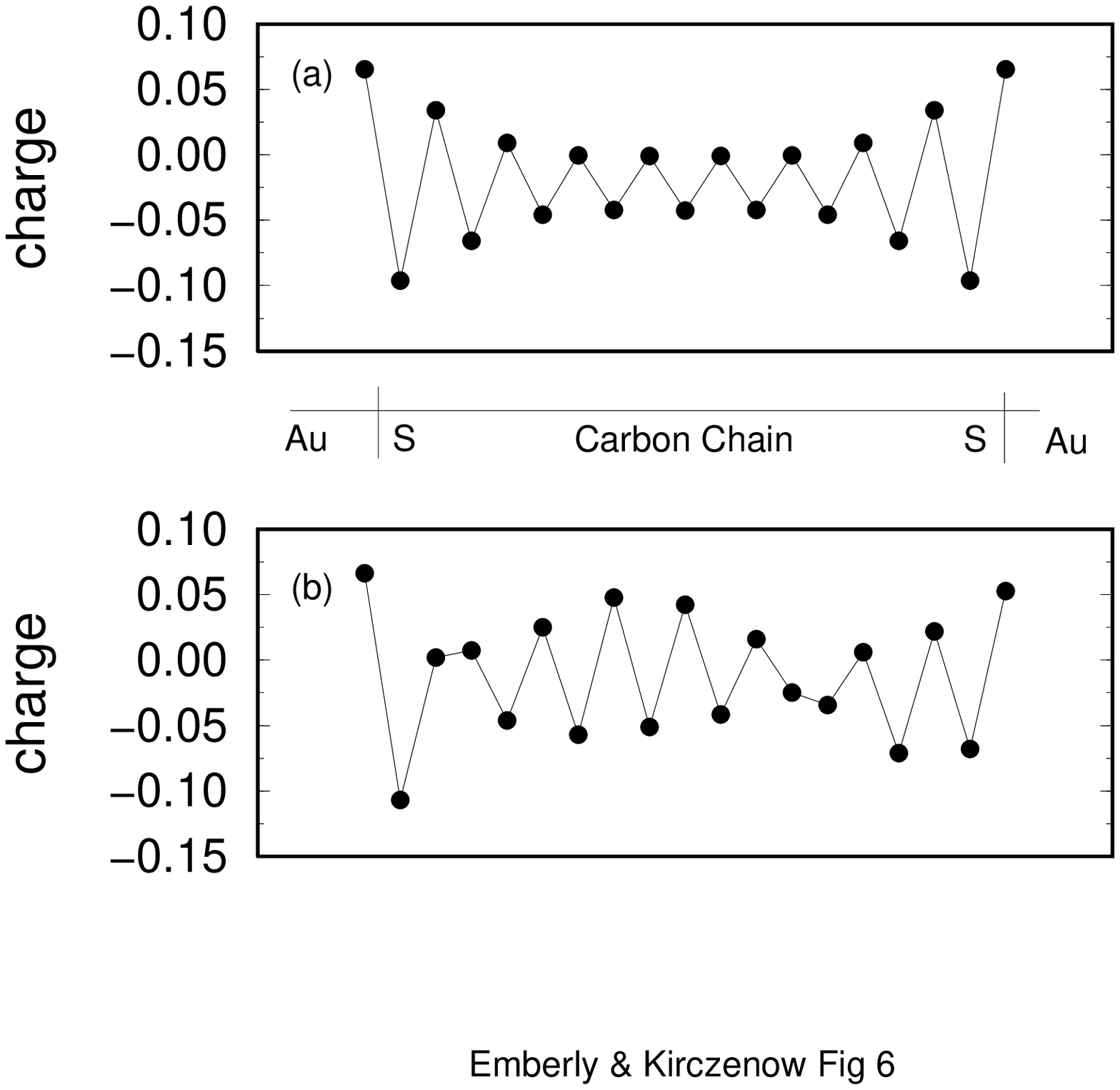}
\caption{Fractional charge along the wire at zero gate bias. (a)
Results for a source-drain bias of 0 V. (b) Results for a source-drain
bias of 1V.}
\label{fig6}
\end{center}
\end{figure}

Examination of the calculated fractional charge along the
molecular chain also reveals some interesting phenomena.  The
results that follow are for the molecular wire with two sulfur
atoms adsorbed on the gate. As mentioned above, undimerized
chains were assumed in the calculation. Fig. \ref{fig6} shows
results for the charge along the wire at zero gate bias. The
plot in (a) corresponds a zero source-drain bias. The first
thing to be noted is the charging at the contacts between the
gold and the sulfurs. The gold loses charge and the sulfur
acquires charge which leads to the formation of the
Schottky-like barrier seen in the previous plots of the
potential. Along the carbon chain a commensurate charge density
wave with a period of 2 atomic spacings is formed. Particularly
interesting are the corresponding results shown in (b) for the
charge along the chain when a 1 V source-drain bias is applied:
The phases of the charge density wave at the two ends of the
carbon chain are pinned by the chemical charge transfer that
occurs there between the gold and sulfur atoms. Thus when the
applied source-drain bias breaks the left-right symmetry along
the chain, the charge redistribution along the chain proceeds
through the formation of a soliton anti-soliton pair along the
wire; the net charge of the soliton near the right end of the
wire is negative and of that on the left is positive as one
might expect intuitively based on the sign of the source-drain
bias. Notice also the associated $\pi$ phase shifts of the
phase of the charge density wave across the soliton and across
the anti-soliton -- this is most easily seen by comparison with
the phase of the charge density wave at zero source-drain bias
shown in Fig. \ref{fig6}(a).

\begin{figure}[!t]
\begin{center}
\includegraphics[bb = 0 0  640 800,clip,width =
0.75\textwidth]{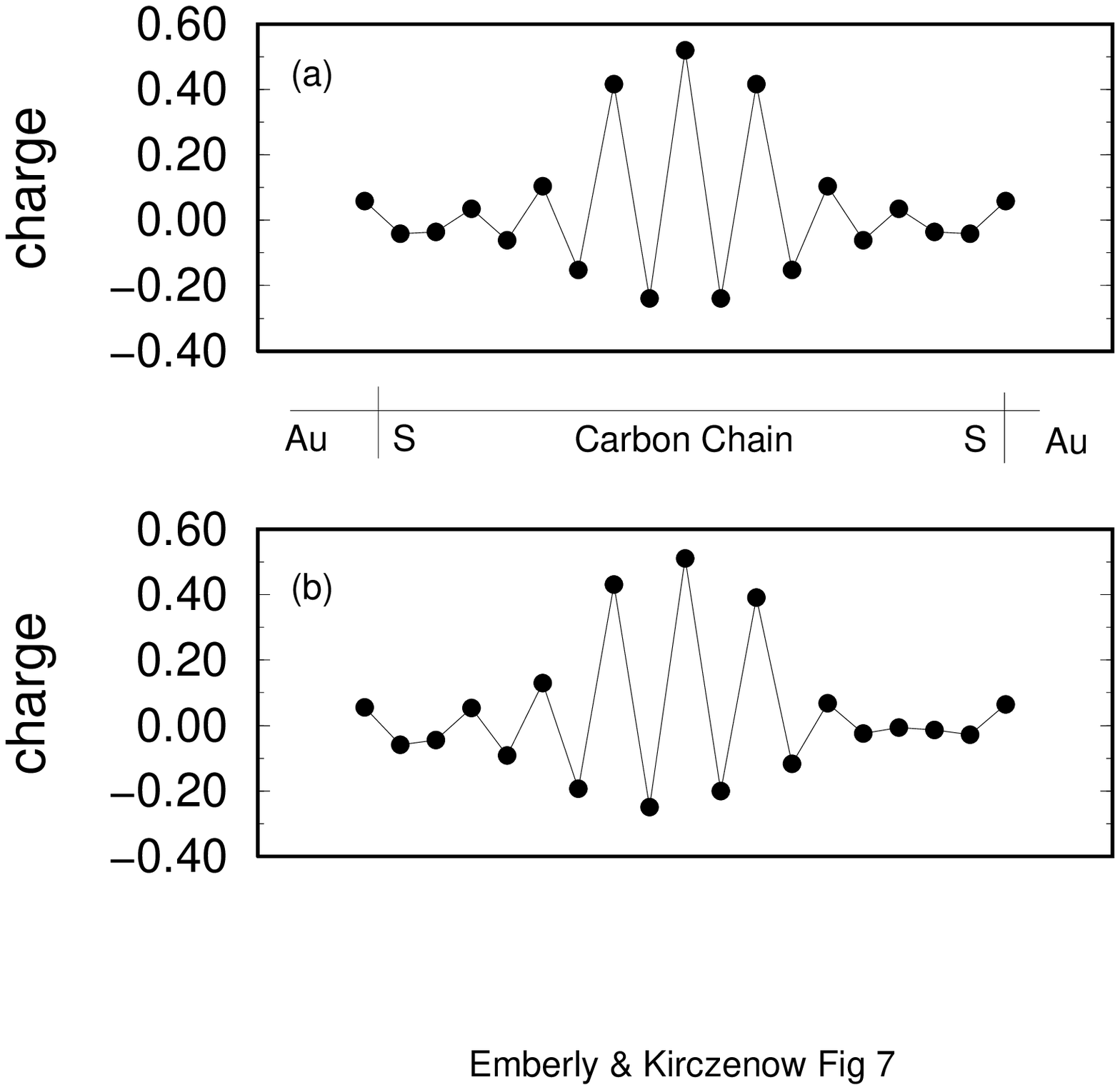}
\caption{Fractional charge along the wire at -3 V gate bias. (a)
Results for a source-drain bias of 0 V. (b) Results for a source-drain
bias of 1V. }
\label{fig7}
\end{center}
\end{figure}

To gain further insight into the role of the solitons in this
system, consider the charge modulation along the wire at a gate
voltage of -3 V that is plotted in Fig. \ref{fig7}. In (a) the
results for a zero source-drain bias are shown. In this case
the phase of the charge density wave at the ends of the chain
is pinned by the chemical charge transfer as before, but over
the gate electrode it is pinned in {\em anti-phase} with the
zero gate voltage case (shown in Fig. \ref{fig6}(a)) by the
interaction with the charged sulfur atoms on the gate. Because
of this there are two symmetric solitons near the ends of the
wire in Fig. \ref{fig7}(a). In contrast to the case shown in
Fig. \ref{fig6}(b), here {\em both} solitons carry a net
negative charge which is a manifestation of the symmetric
redistribution of the electronic charge density from the center
of the wire towards its ends in response to the negative bias
voltage applied the gate.

When a 1~V source-drain bias is applied to this system (Fig.
\ref{fig7}(b)) another soliton appears to be forming near the
right end of the wire between the last two carbon atoms, again
breaking the left-right symmetry of the charge distribution in
response to the applied source-drain voltage. Thus the system
responds to the application of moderate gate or source-drain
voltages through the formation of additional solitons along the
wire.

Further investigation of the dynamics and energetics of the
formation of solitons in these carbon chains and of the
underlying physical mechanisms will clearly be of interest.

\section{Conclusions}
In summary, we have provided a theoretical framework for
modelling current flow and charging effects in multi-terminal
molecular wire systems. Included in our methodology is a
procedure for evaluating the self-consistent electro-static
potential in the region of the molecule due to application of
voltages on the different terminals. It is based on an
extension of tight-binding theory to allow for fractional
charging. Also taken into consideration is the induced charge
on the metallic leads that is handled by means of image charge
potentials.

This theory was applied to the simulation of a molecular wire
transistor (MWT) that can be realized using presently available
experimental techniques.  It was shown that it is feasible to
control the current flowing through this device by the
application of a gate voltage. The key to controlling the
current is the use of a suitable adsorbate on the gate
electrode that is effective in mediating the gate voltage onto
the molecular wire. It was also shown that charged solitons
form along the wire in response to the application of voltages
to the terminals. Further research into the dynamics of these
solitons will be of interest. The incorporation of a third
terminal into molecular wire systems as described in this work
should open up many exciting possibilities in the science and
technology of molecular nano-electronic devices.

We thank R. Hill, M. A. Reed, H. Guo, B. Heinrich, M. Freeman,
J. Young and M. Moskovits for rewarding discussions. This work
was supported by NSERC and by the Canadian Institute for
Advanced Research.

\section{Appendix:  Potential for Three Terminal Geometry}
\begin{figure}[!t]
\begin{center}
\includegraphics[bb = 0 0  640 800,clip,width =
0.75\textwidth]{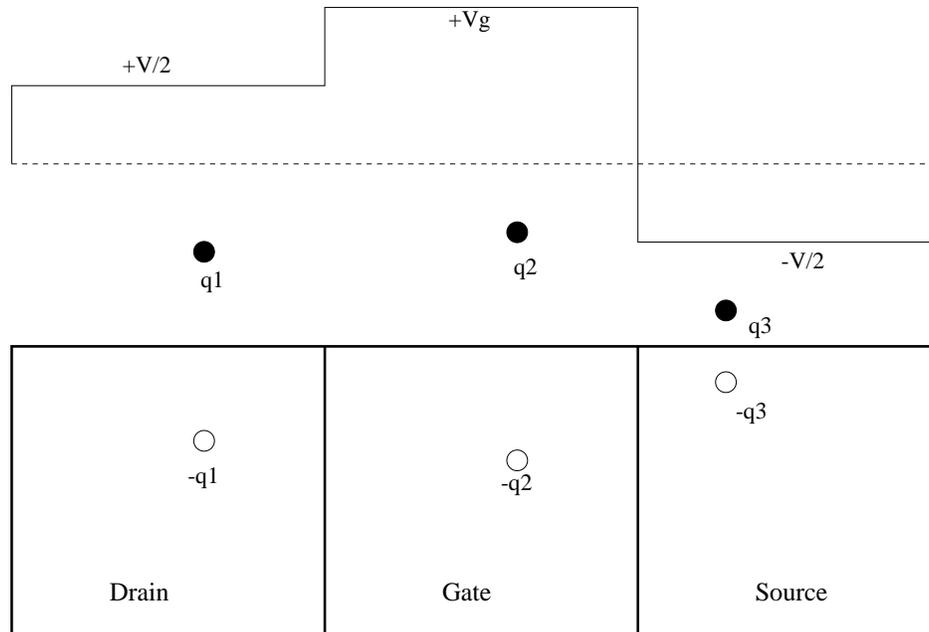}
\caption[A schematic of the three terminal electro-static problem]{A
schematic of the three terminal electro-static problem. The metallic
substrate is divided into three equal sized regions that act as the
source, the gate and the drain. The source is held at constant
potential of $-V/2$, the gate at $V_g$ and the drain at $V/2$. A
charge $q$ above the metallic conducting substrate generates an image
charge at an equal distance below the surface of the conductor. }
\label{fig8}
\end{center}
\end{figure}

This appendix evaluates the electrostatic potential associated
with the terminals and image charges that is used in the
simulation presented in Sec.~V. The geometry that is considered
here is a planar three terminal configuration (this is shown in
Fig.~(\ref{fig8}). It consists of a substrate divided into
three regions. The first region from $-3L/2<x<-L/2$ acts as the
source, the second region from $-L/2<x<L/2$ acts as the gate
and the third region from $L/2<x<3L/2$ is the drain. (In
reality such a substrate would have insulating dielectrics
between the three regions, but that is not considered
explicitly here). The distance perpendicular to the substrate
is measured by the co-ordinate $z$. The potential
$\phi(\vec{r})$ satisfies Laplace's equation,
\begin{equation}
\frac{\partial^2 \Phi(x,z)}{\partial z^2} + \frac{\partial^2
\Phi(x,z)}{\partial x^2} = 0
\end{equation}
subject to the boundary conditions, $\phi(x,0) = -V/2$ for
$-3L/2<x<-L/2$, $\phi(x,0) = V_g$ for $-L/2<x<L/2$ and
$\phi(x,0)= V/2$ for $L/2<x<3L/2$.

The above equation has the solution,
\begin{equation}
\Phi(x,z) = \int_{-\infty}^{\infty} d\alpha \, e^{i \alpha x}
e^{-\alpha z} c(\alpha) \label{eq:Phi_xz}
\end{equation}
Applying the boundary condition at $z= 0$ gives,
\begin{equation}
\Phi(x,0) = \int_{-\infty}^{\infty} d\alpha \, e^{i \alpha x} c(\alpha)
\end{equation}
Performing the inverse transform allows the solution for
$c(\alpha)$,
\begin{equation}
c(\alpha) = \frac{1}{2\pi} \int_{-\infty}^{\infty} dx \, \Phi(x,0) e^{-i
\alpha x}
\end{equation}
With the boundary conditions above, it is found that this gives
\begin{equation}
c(\alpha) = \frac{1}{2\pi}\left\{ \frac{V}{i \alpha}[
\cos(\frac{3L\alpha}{2}) - \cos(\frac{L\alpha}{2}) ] + \frac{2
V_g}{\alpha} \sin(\frac{L\alpha}{2}) \right \}
\end{equation}
Inserting this back into Eq.~(\ref{eq:Phi_xz}) and integrating
gives the following expression for the potential,
\begin{eqnarray}
\Phi(x,z) & = & \frac{1}{2\pi} \left \{ -V [\arctan(\frac{3L/2+x}{z}) -
\arctan(\frac{3L/2-x}{z}) \right . \nonumber \\  & & + \left .
  \arctan(\frac{L/2-x}{z}) - \arctan({L/2+x}{z})] \right . \nonumber \\
  & & + \left . 2 V_g [\arctan(\frac{L/2+x}{z}) -\arctan(\frac{x-L/2}{z})]
  \right \}
\end{eqnarray}

The last problem that remains is to find the image potential
arising from the presence of charge $\{q_j\}$ on the
molecule. The substrate is assumed to form a perfect planar
conductor, and hence the method of images applies. For a
charge, $q_j$ at $\vec{r}_j=(x_j,y_j,z_j)$, there is mirror
image charge $-q_j$ at $(x_j,-y_j,z_j)$. Thus the image
potential at position $\vec{r}=(x,y,z)$ is
\begin{equation}
(\phi_{image}(\vec{r},\vec{r}_j))_j = \frac{-q_j}{4\pi \epsilon_0}
\frac{1}{\sqrt{(x-x_j)^2+(y+y_j)^2+(z-z_j)^2}}
\end{equation}
The total image potential is found by summing over image
potentials arising from all the charges, $\{q_j\}$ on the
molecule


\end{document}